\documentclass[preprint,aps,amsmath,amssymb,showpacs]{revtex4}

\usepackage{graphicx}% Include figure files
\usepackage{dcolumn}% Align table columns on decimal point
\usepackage{bm}% bold math

\begin{document}

\title{Vertical Confinement and Evolution of Reentrant Insulating 
Transition in the Fractional Quantum Hall Regime}
% \title{Reentrant Insulating Phase near $\nu = 1/3$ Fractional Quantum 
% Hall Effect in a Vertically Confined Two-Dimensional Electron System}

\author{I. Yang$^{1}$, W. Kang$^{1}$, S.T. Hannahs$^{2}$, L.N. Pfeiffer$^{3}$, 
and K.W. West$^{3}$}

\affiliation{
$^{1}$James Franck Institute and Department of Physics,
 University of Chicago, Chicago, Illinois 60637\\
 $^{2}$National High Magnetic Field Laboratory, 1800 East Paul 
Dirac Drive, Tallahassee, Florida 32310\\
$^{3}$Bell Laboratories, Lucent Technologies, 600 Mountain
Avenue, Murray Hill, New Jersey 07974}

\date{\today}

\begin{abstract}
We have observed an anomalous shift of the high field reentrant insulating 
phases in a two-dimensional electron system (2DES) tightly confined 
within a narrow GaAs/AlGaAs quantum well. Instead of the well-known
transitions into the high field insulating states centered around 
$\nu = 1/5$, the 2DES confined within an 80\AA-wide quantum well exhibits 
the transition at $\nu = 1/3$. Comparably large quantum lifetime of the 2DES 
in narrow well discounts the effect of disorder and points to 
confinement as the primary driving force behind the evolution of the 
reentrant transition.
\end{abstract}


\maketitle

The prospect for a quantum Wigner crystal has driven the study of
two-dimensional electron system (2DES) under the conditions of 
low temperature and high magnetic field\cite{WignerCrystal}.  
Experiments on high mobility GaAs/AlGaAs heterostructures have 
shown that in the limit of zero temperature the $\nu = p/(2p \pm 1), 
p = 1, 2, 3\ldots$ series of incompressible quantum liquid states of 
fractional quantum Hall effect (FQHE) terminates with a transition 
into a field-induced insulator at low fillings. A dramatic sequence 
of transitions involving a reentrant insulating phase, a $\nu = 1/m$ 
primary FQHE state, and the final insulating phase can be realized 
in the highest quality samples.  
Initially observed in 2DES around the $\nu = 1/5$ FQHE\cite{Jiang90}, the 
reentrant insulating phase has been also detected adjacent to the 
$\nu = 1/3$ FQHE state in two-dimensional hole system (2DHS)\cite{Santos92}. 
Reentrant insulating behavior is also seen in high quality
silicon MOSFET and $p$-SiGe heterostructures 
in the integer quantum Hall regime\cite{D'Iorio92,Kravchenko95,Sakr01}.
Because of the correlated nature of the FQHE, the insulating states 
above and below the FQHE states at $\nu = 1/3$ and 1/5 are expected 
to be driven by electron-electron correlation rather than disorder, 
enhancing the likelihood of formation of a Wigner crystal. This has 
given impetus for various electrical, acoustic, microwave, and optical
investigations of the high field insulating 
phases\cite{Goldman90,Li91,Williams91,Buhmann91,Paalanen92,Goldys92,Manoharan94,Li97,Pan02}. 

For an ideal, disorder-free 2DES, the ground state in the limit of zero 
temperature is an ordered electron crystal at small Landau level fillings. 
However, a positive confirmation of the Wigner 
crystalline order has remained controversial and somewhat elusive. This 
is partly due to absence of scattering experiments that can directly probe 
the crystalline order of the insulating phases. In addition, 
interpretation of various experiments in the Wigner crystalline regime 
is complicated by general lack of understanding of the effects of 
disorder in presence of strong interaction. 
Even the highest quality samples presently available possess 
non-negligible disorder at low fillings, and the competition of 
disorder and interaction is thought to modify the ground state 
of 2D systems in some fundamental way. At short distances, the 
ground state is conjectured to evolve into a partially ordered Wigner 
crystal consisting of finite-size domains that are pinned by the disorder 
potential\cite{WignerCrystal}. In presence of strong disorder, the ground 
state evolves into a disorder-driven correlated insulator called 
Hall insulator\cite{Kivelson92,Shahar95,Wong96}.

In this paper we present an unexpected observation of reentrant 
insulating phase around the $\nu = 1/3$ FQHE state in a two-dimensional 
{\it electron} system. The 2DES in question is found in 
a GaAs/AlGaAs quantum well whose narrow width produces a tight 
vertical confinement of the electronic wave function. 
In spite of its relatively low mobility, we observe a clear sequence 
transitions to an insulator, a FQHE, and back to an insulating phase 
in the vicinity of $\nu = 1/3$ filling in a fashion reminiscent of the 
Wigner crystalline regime in high mobility $n-$ and $p$-type 
GaAs/AlGaAs heterostructure. 
Tilted magnetic field study shows that the insulating phases are found 
to be insensitive to the presence of parallel magnetic field. We also
compare the properties of the 2DES in the NQW with a wider quantum well 
and the conventional heterostructure. We find that the single particle 
lifetime of the 2DES in NQW is comparable to other higher mobility specimen 
and conclude that confinement within the NQW may be responsible for the 
shift in the reentrant transition.

The experiment was performed using a modulation doped AlGaAs/GaAs 
quantum well of 80-\AA\ in width. The density of the sample was $n = 
1.1 \times 10^{11} \rm{cm}^{-2}$ with a low temperature mobility of $2.56 
\times 10^{5} \rm{cm}^{2}/\rm{V~sec}.$  The transport in narrow 
quantum well (NQW) is dominated by interfacial fluctuations\cite{Motohisa92}, 
yielding 
a substantially lower mobility than comparable heterostructure samples.
Samples in Hall bar and van der Pauw configurations were studied 
inside a dilution refrigerator with a 14 tesla superconducting magnet.
Tilted-field study was performed up to 40 tesla using the 
hybrid magnet at the National High Magnetic Field Laboratory. 
A light emitting diode
was used to illuminate the sample at low temperatures. 
Depending on the illumination conditions 
and the thermal cycling history, small variation in the data was detected.
This, however, does not alter our conclusions.

Fig. \ref{fig:narrow1} illustrates the magnetoresistance $R_{xx}$ of 
a NQW sample at a
temperature of 35 mK. The most striking feature of the data is the 
sharp increase of longitudinal resistance between $\frac{1}{3} < \nu < 0.45$. 
Slightly after $\nu = 1/2$ $R_{xx}$ increases dramatically by more 
than 2 orders of magnitude, completely overwhelming other transport 
features. The peak resistance at $\nu = 0.38$ exceeds $900 \rm{k}\Omega$ 
before dropping precipitously as it enters the $\nu = 1/3$ FQHE 
state. $R_{xx}$ subsequently diverges upon entering a high field
insulating phase. The inset of Fig. \ref{fig:narrow1} illustrates
longitudinal and Hall resistances of the same sample under a 
different illumination condition. The $R_{xy}$ is quantized at 
$3h/e^{2}$ at $\nu = 1/3$, demonstrating the formation of the 
$\nu = 1/3$ FQHE state. The $\nu = 3/7$ 
FQHE state is found as a weak $R_{xx}$ minimum prior to the 
reentrant insulating phase.  

Fig. \ref{fig:narrow2} shows the temperature dependence of the 
insulating state at a 
slightly higher density ($n = 1.2 \times 10^{11} \rm{cm}^{-2}$). As 
temperature is raised, $R_{xx}$ decreases sharply with the insulating 
features largely disappearing above T = 300 mK.
The inset of Fig. \ref{fig:narrow2} 
shows an Arrhenius plot of peak resistance at $\nu = 
0.38$. The resistance at the peak is activated with an activation 
energy of $E_{g} \sim 0.26 $K. 
At lower temperatures, there is a
saturation of the resistivity. Measurement of I-V characteristics 
shows that the transport in the insulating regime is highly 
nonlinear, similar to previously observed
reentrant insulating 
phases\cite{Jiang90,Santos92,Goldman90,Li91,Williams91,D'Iorio92,Kravchenko95,Sakr01}. 

Fig. \ref{fig:narrow3} shows the effect of parallel magnetic field 
in the insulating regime above and below the $\nu = 1/3$ FQHE state.
Comparison of magnetoresistance for the tilt angle of $\theta = 
0^{\circ}$ and $\theta = 58^{\circ}$ at T = 50 mK shows that there 
is no appreciable change in both $\nu = 1/3$ FQHE state and the 
insulating states above and below $\nu = 1/3$ even though the total 
magnetic field was nearly double of the perpendicular magnetic field. 
This shows that neither the increase in the Zeeman energy nor the 
deformation of the wave function due to strong parallel magnetic field
appears to play a large role in the insulating phase.

In Fig. \ref{fig:narrow4} we explore the role of vertical confinement 
by comparing the transport between quantum well samples with different 
widths. Fig. \ref{fig:narrow4}a illustrates magnetoresistance of our
80\AA\ wide NQW. Fig. \ref{fig:narrow4}b illustrates the 
magnetoresistance of a quantum well sample that is 300\AA\ wide and 
possessing a mobility of $\mu = 7.8 \times 10^{6} cm^{2}/Vs$ and 
a density of $n = 6.2 \times 10^{10} cm^{-2}.$ In the wider quantum 
well specimen, magnetoresistance at 30 mK shows 
a well-developed sequence of FQHE states centered around  $\nu = 1/2$ 
and $\nu = 1/4$ followed by the reentrant insulating phase prior to the
$\nu = 1/5$ FQHE state. This contrasts sharply with the 80 \AA\ -wide 
NQW which only exhibits a weakly developed FQHE state at $\nu = 2/3$ 
prior to the reentrant insulating state above $\nu = 1/3$. These 
results suggest strongly that narrow confinement is likely to be
important in altering the properties of 2DES in the insulating regime.

The prevailing view on the high field reentrant transitions in 
various 2DES involves 
either an entry into the Wigner crystalline regime\cite{WignerCrystal} or 
an approach based on the global phase diagram of quantum Hall 
effect\cite{Kivelson92}. Consideration of the former scenario follows 
from the importance of the electron-electron interaction in the limit 
of low fillings. Since the Laughlin states at $\nu = 1/m, m = 3, 5...$ 
occur from strong 
electronic correlation\cite{Laughlin83}, it follows that the interaction 
should also play an important role in the adjacent insulating states. 
The reentrance is explained in terms of competition between the FQHE 
liquid and Wigner solid. In this point of view, the insulating phases 
seen in the NQW is likely to 
be some kind of strongly correlated ground state driven by interaction.
Alternatively, the insulating phases in NQW may occur due to some electron
localization effect associated with disorder as suggested by its modest 
mobility. In this context, the global phase diagram picture of quantum 
Hall transitions\cite{Kivelson92} is relevant as a transition
from an insulator into a FQHE state at $\nu = 1/3$ is permitted.

However, in either pictures, there is no obvious explanation for 
the shift in the reentrant insulating transition to $\nu = 1/3$ for 
the 2DES in the NQW. This feature is particularly puzzling since 
the reentrant behavior is always found near $\nu = 1/5$ in 2DES based 
on GaAs/AlGaAs structures\cite{WignerCrystal,Jiang90}. In fact, the observed sequence 
of transitions in NQW resembles the reentrant insulating transitions near 
$\nu = 1/3$ in 2DHS in GaAs/AlGaAs heterostructure\cite{Santos92}. 
In the reentrant insulating transitions seen in Si MOSFET and p-SiGe 
heterostructure, the reentrant insulating phases are found in the integer 
quantum Hall regime and spin is thought to play an important 
role\cite{D'Iorio92,Kravchenko95,Sakr01}. In our NQW, tilted field experiment
appears to rule out the role of spin in the reentrant insulating phase. 

Theoretically the transition into a Wigner crystal in 2DES 
is predicted for $\nu \leq 1/6.5$\cite{Lam84,Levesque84,Zhu95}. 
In the case of 2DHS in GaAs/AlGaAs heterostructures, the stability of the
insulating phases near $\nu = 1/3$ is explained in terms of increased Landau 
level mixing associated with its heavier effective mass 
($m^{*}_{h} = 0.3m_{\circ}$) compared to that of electrons 
($m^{*}_{e} = 0.067m_{\circ}$)\cite{Santos92,Zhu93,Price93}. 
Since the effective mass of electrons in narrow quantum wells has
been shown to be comparable to heterostructures\cite{Huant92}, 
Landau level mixing does not appear to play a significant role in NQW. 
As there is no universally accepted explanation for the reentrant transitions 
observed in various two-dimensional semiconductor systems, understanding 
of the shift in the reentrant behavior in the NQW is likely to be important 
in clarifying the nature of the associated insulating phases.

The role of disorder remains an important 
question as the mobility of 2DES in NQW is modest compared to other 
2DES based on GaAs/AlGaAs quantum structures. 
However, previous experiments on GaAs/AlGaAs heterostructures with mobilities 
comparable to that of our NQW have not found reentrant behavior next to 
the  $\nu = 1/3$ FQHE state\cite{Shahar95,Wong96}. While this 
appears to discount the importance of disorder associated with the 
reentrant behavior in the NQW, interfacial roughness serves to 
restrict the transport in narrow quantum wells\cite{Motohisa92} and 
a more quantitative measure of disorder is necessary. This is 
particularly important since electronic transport under magnetic 
field is largely determined by large angle scattering instead of the small 
angle scattering which dominates the zero field transport. The single 
particle relaxation time, $\tau_{s}$, in semiconductors is consequently
substantially smaller than the transport scattering time, 
$\tau_{t}$\cite{DasSarma85,Coleridge91}. 

In Table \ref{tab:table1}  we summarize the 
properties of 2DES derived from 3 different GaAs/AlGaAs structures that 
exhibit reentrant insulating phase in the lowest Landau level. In addition 
to the 80\AA\ quantum well, 2DES from a heterostructure and 300\AA\ quantum 
well with mobilities that are respectively 10 and 30 times larger were 
compared. $\tau_{t}$ was deduced from zero field mobility, $\mu = 
e\tau_{t}/m^{*}$, and $\tau_{s}$ was determined from the Dingle analysis 
of the Shubnikov-de Haas (SDH) oscillations as suggested by 
Coleridge\cite{Coleridge91}. In contrast to the large differences 
in $\tau_{t}$, we found that $\tau_{s}$'s determined from different structures
were surprisingly close. The $\tau_{s}$ in  the high mobility 
300 \AA\ wide quantum well was 8.5 $ps$ with the 80 \AA\ NQW yielding
a $\tau_{s}$ of 3.7 $ps$. The heterostructure sample was
 found to possess a $\tau_{s}$ comparable to other 2DES samples.
 
The proximity of $\tau_{s}$'s is also reflected in the onset of SDH oscillations, $B_{onset}$. In 
the insets of Fig. \ref{fig:narrow4} we show the SDH oscillations for 
80 \AA\ NQW and 300 \AA\ quantum well. A $B_{onset}$ of $\sim$ 60 mT 
for the NQW and $\sim 30$mT for other samples were obtained. The
comparable $\tau_{s}$ and $B_{onset}$ suggest that 
in spite of its low mobility, the electronic lifetime in NQW is not 
adversely affected by the tight confinement, increasing the likelihood
that the reentrant behavior around $\nu = 1/3$ is driven by 
interaction rather than disorder. On the other hand, the larger 
resistivity at $\nu = 1/2$ and the absence of high order FQHE states 
is consistent with presence of stronger disorder in the NQW than the 
wider well. However, coexistence of the $\nu = 1/3$ FQHE state next 
to the insulating phases indicates that disorder is not enough to 
suppress the electron correlation in the limit of low fillings.

Since the most distinguishing characteristic of the 2DES in a NQW involves its 
vertical confinement, the physics of reentrance may potentially occur from 
a confinement-induced evolution of the Coulomb interaction. The ground state 
energy and the interaction parameters of 2DES depends on the finite vertical 
extent of the electronic wave function\cite{Zhang86,Morf02}. A thicker 2DES 
consequently experiences a softer 
Coulomb potential and thereby possesses a reduced FQHE energy gap compared to a 
thinner 2DES. Our estimate of the thickness of electrons based on the interaction
parameters in quantum wells\cite{DasSarma85a}
 points to a substantial reduction in its thickness
compared to the 2DES found in heterostructures\cite{Morf02}. For the 2DES in
Table \ref{tab:table1}, we obtain a thickness of 128\AA\ in the heterostructure specimen
which contrasts sharply against the estimated thickness of 19\AA\ and 67\AA\ for 
the 80\AA\ and 300\AA\ quantum wells. The role of thickness in the insulating 
phases in 2DES remains unknown and further theoretical investigation is 
necessary to clarify the effect of confinement in relation to the enhancement 
of $r_{s}$ in the insulating regime.

In summary, we have observed a puzzling shift in the reentrant insulating 
phases in a 2DES confined within a NQW. The apparent shift in the reentrant 
transition to $\nu = 1/3$ cannot be reconciled in terms of disorder and points
to importance of confinement. Understanding the shift in the reentrant transition 
may be important in uncovering the physics behind the reentrant insulating
phases in the NQW as well as other two-dimensional electron and hole systems.
While the confinement within a NQW is expected to produce a thinner 2DES, its 
effect in the high field insulating states remains to be clarified.

We would like to thank H. Fertig, H.W. Jiang, A.H. MacDonald, H.L. 
Stormer, and D.C. Tsui for useful 
discussions. A portion of this work was performed at the National 
High Magnetic Field Laboratory which 
is supported by NSF Cooperative Agreement No. DMR-9527035.
The work at the University of Chicago is supported by NSF 
DMR-9808595 and DMR-0203679.

\pagebreak

\pagebreak

\begin{table*}
\caption{\label{tab:table1} Comparison of the 
properties of two-dimensional electron systems in 80\AA\ GaAs/AlGaAs 
quantum well, 300\AA\ GaAs/AlGaAs quantum well, and a GaAs/AlGaAs
heterostructures. All exhibit reentrant behavior in 
the lowest Landau level.}
\begin{ruledtabular}
\begin{tabular}{cccccccccc}
% \begin{tabular}{c.cc.c} \hline\hline
sample & density & mobility, $\mu$ & scattering 
& onset of & single particle & $\tau_{t}/\tau_{s}$ & $\rho_{xx}$ 
at $\nu = 1/2$  & reentrance \\
  & ($10^{11} {cm}^{-2}$) & ($\rm{cm}^{2}/\rm{Vs}$) & time, $\tau_{t}$$(ps)$ & 
  SDH (mT) & lifetime, $\tau_{s}$$(ps)$ & & ($k\Omega/\Box$) & \\
\hline
% \vspace*{1mm}
%     &&&&& 
80\AA\ QW & 1.1 & $2.5 \times 10^{5}$ & 9.7 & 61 & 3.6 & 2.7 &  15.3 & $\nu = 1/3$ \\  
% heterostructure & 1.1 & $6.5 \times 10^{6}$ & 248 & 46 & 5.5 & 45 & $\nu = 1/5$ \\
300\AA\ QW & 0.62 & $7.8 \times 10^{6}$ & 297 & 29 & 8.5 & 35 & 0.241 & $\nu = 1/5$ \\
heterostructure & 0.53 & $2.5 \times 10^{6}$ & 95 & 30 & 6.7 & 
14 & 0.741 & $\nu = 1/5$ \\
% \hline\hline

\end{tabular}
\end{ruledtabular}
\end{table*}

\vspace*{4in}
\pagebreak

\begin{figure}
% \vspace*{1.5in}
%\includegraphics[width=3.25in]{nqwfig1.eps}
\caption{\label{fig:narrow1} 
Longitudinal magnetoresistance of a narrow quantum well at 35 mK. 
Inset: longitudinal (blue) and transverse (red) magnetoresistivities of a sample 
with a slightly higher density. Integers and fractions indicate the Landau level 
filling of the two-dimensional electron system.}
\end{figure}

\begin{figure}
% \vspace*{1.25in}
%\includegraphics[width=3.25in]{nqwfig2.eps}
\caption{\label{fig:narrow2} Longitudinal magnetoresistance of a narrow 
quantum well at 32, 46, 60, 80, 100, 140, 210, 300 and 420 mK. Inset: 
Arrhenius plot of the 
resistivity at the peak of the reentrant insulating phase.}
\end{figure}

\begin{figure}
% \vspace*{1.25in}
%\includegraphics[width=3.25in]{nqwfig3.eps}
\caption{\label{fig:narrow3} Magnetoresistance vs perpendicular 
magnetic field for tilt angles at $\theta = 0^{\circ}$ (solid) and 
$\theta = 58^{\circ}$ (dashed). The measurement was performed at 
50 mK.}
\end{figure}

\begin{figure}
% \vspace*{1.25in}
%\includegraphics[width=3.25in]{nqwfig4.eps}
\caption{\label{fig:narrow4} Comparison of reentrant insulating 
transitions in a  80 \AA\ (top) and 300 \AA\ (bottom) wide quantum well with 
mobility of $2.5\times 10^5 cm^{2}/Vs$ and $7.8\times 10^{6}cm^{2}/Vs$, 
respectively, at T $\approx 30mK$.  
Insets: expanded view of the low field Shubnikov-de Haas (SDH)
oscillations. While the mobilities differ by a factor of $\sim$30, 
the onsets of SDH oscillations differ by a factor of $\sim$2.}
\end{figure}


\begin{thebibliography}{99}
\bibitem{WignerCrystal} See for example, review articles by H.A. Fertig and 
 M. Shayegan, in {\it Perspectives in Quantum Hall Effects,} edited by S. 
 Das Sarma and A. Pinczuk (Wiley, New York, 1996), and references therein. 
\bibitem{Jiang90} H.W. Jiang {\it et al.}, Phys. Rev. Lett. {\bf 65}, 633 (1990). 
\bibitem{Santos92} M.B. Santos {\it et al.}, Phys. Rev. Lett. {\bf  68}, 1188 (1992). 
\bibitem{D'Iorio92} M.D. D'Iorio {\it et al.}, Phys. Rev. B. {\bf 46}, 15992 (1992).
\bibitem{Kravchenko95} S. Kravchenko {\it et al.}, Phys. Rev. Lett. {\bf 75}, 910 (1995).
\bibitem{Sakr01} M.R. Sakr {\it et al.}, Phys. Rev. B {\bf 64}, 161308R (2001).
\bibitem{Goldman90} V.J. Goldman {\it et al.}, Phys. Rev. Lett. {\bf 65} 2189 (1990). 
\bibitem{Li91} Y.P. Li {\it et al.}, Phys. Rev. Lett. {\bf 67}, 1630 (1991).  
\bibitem{Williams91} F.I.B. Williams {\it et al.}, Phys. Rev. Lett. {\bf 66}, 3285 (1991). 
\bibitem{Buhmann91} H. Buhmann {\it et al.}, Phys. Rev. Lett. {\bf 66}, 926 (1991). 
\bibitem{Paalanen92} M.A. Paalanen {\it et al.}, Phys. Rev. B {\bf 45}, 11342 (1992). 
\bibitem{Goldys92} E.M. Goldys {\it et al.}, Phys. Rev.  B {\bf 46}, 7957 (1992). 
\bibitem{Manoharan94} H.C. Manoharan {\it et al}, Phys. Rev. B {\bf 50}, 17662 (1994).
\bibitem{Li97} C.C. Li {\it et al.}, Phys. Rev. Lett. {\bf 79}, 1353 (1997).  
\bibitem{Pan02} W. Pan {\it et al.}, Phys. Rev. Lett. {\bf 88}, 176802 (2002). 
\bibitem{Kivelson92} S. Kivelson {\it et al}, Phys. Rev. B {\bf 46}, 2223 (1992). 
\bibitem{Shahar95} D. Shahar {\it et al.}, Phys. Rev. Lett. {\bf 74}, 4511 (1995).  
\bibitem{Wong96} L.W. Wong {\it et al}, Phys. Rev. B {\bf 54}, 17323 (1996). 
\bibitem{Motohisa92} J. Motohisa {\it et al}, Appl. Phys. Lett. {\bf 60}, 1315 (1992).
\bibitem{Lam84} P.K. Lam and S.M. Girvin, Phys. Rev. B {\bf 30}, 473 (1984). 
\bibitem{Levesque84} D. Levesque {\it et al}, Phys. Rev. 
 B {\bf 30}, 1056 (1984). 
\bibitem{Zhu95} X. Zhu and S.G. Louie, Phys. Rev. B {\bf 52}, 5863 (1995). 
\bibitem{Zhu93} X. Zhu and S.G. Louie, Phys. Rev. Lett. {\bf 70}, 335 (1993). 
\bibitem{Price93} R. Price {\it et al}, Phys. Rev. Lett. 
 {\bf 70}, 339 (1993). 
\bibitem{Huant92} S. Huant {\it et al}, Phys. Rev. B {\bf 46}, 2613 (1989). 
\bibitem{Laughlin83} R.B. Laughlin, Phys. Rev. Lett {\bf 50}, 
1395 (1983).
\bibitem{DasSarma85} S. Das Sarma and F. Stern, Phys. Rev. B {\bf 32}, 8442 (1985).
\bibitem{Coleridge91} P.T. Coleridge, Phys. Rev. B {\bf 44}, 3793 (1991).
\bibitem{Zhang86} F.C. Zhang  {\it et al}, Phys. Rev. B {\bf 33}, 2903 (1986).
\bibitem{Morf02} R.H. Morf {\it et al}, Phys. Rev. B {\bf 66}, 075408 (2002).
\bibitem{Ortalano97} M.W. Ortalano {\it et al}, Phys. Rev. B {\bf 55}, 7702 (1997).
\bibitem{DasSarma85a} S. Das Sarma {\it et al}, Ann. Phys. (N.Y.) {\bf 163}, 78 (1985).
\end{thebibliography}
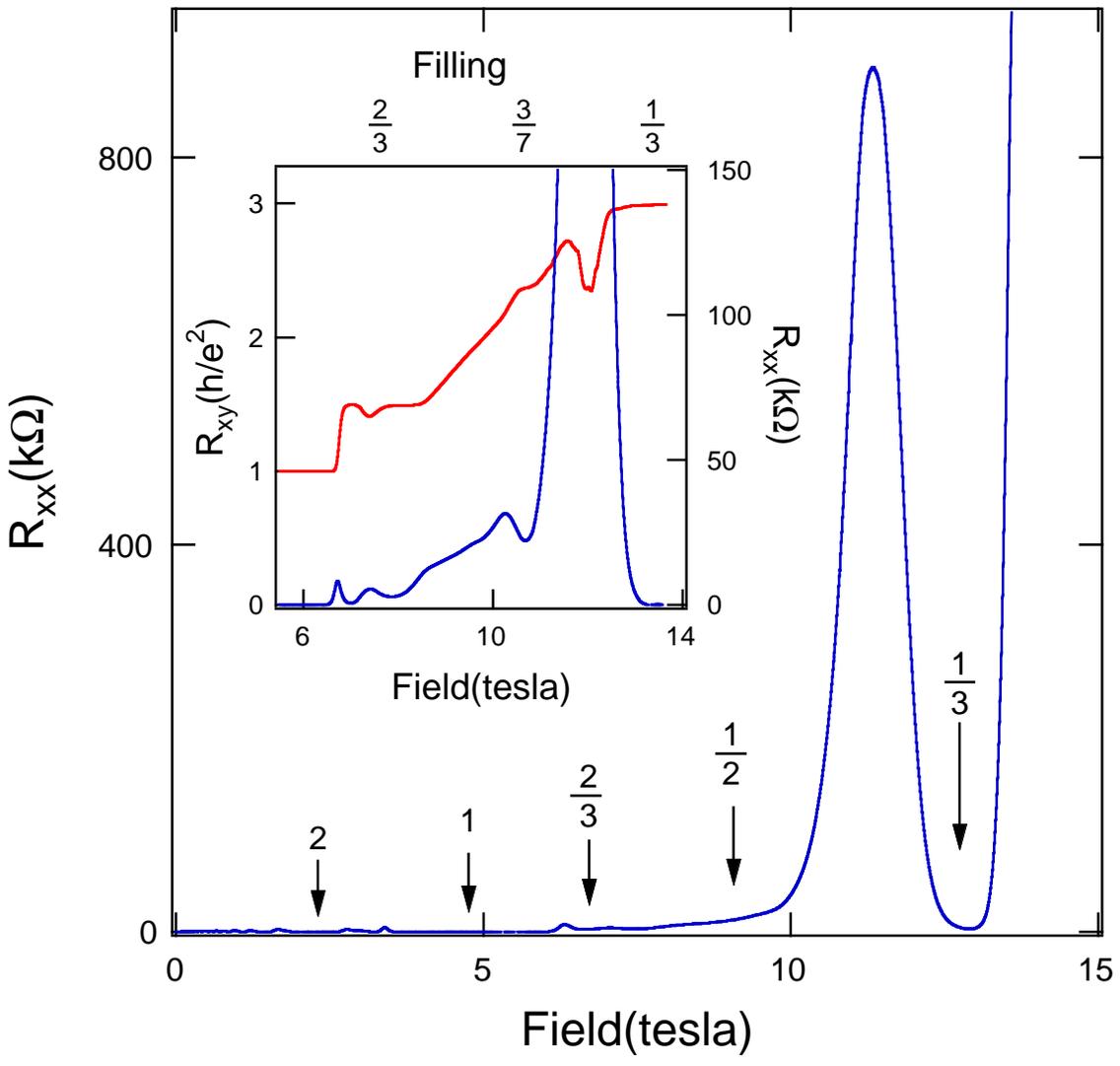
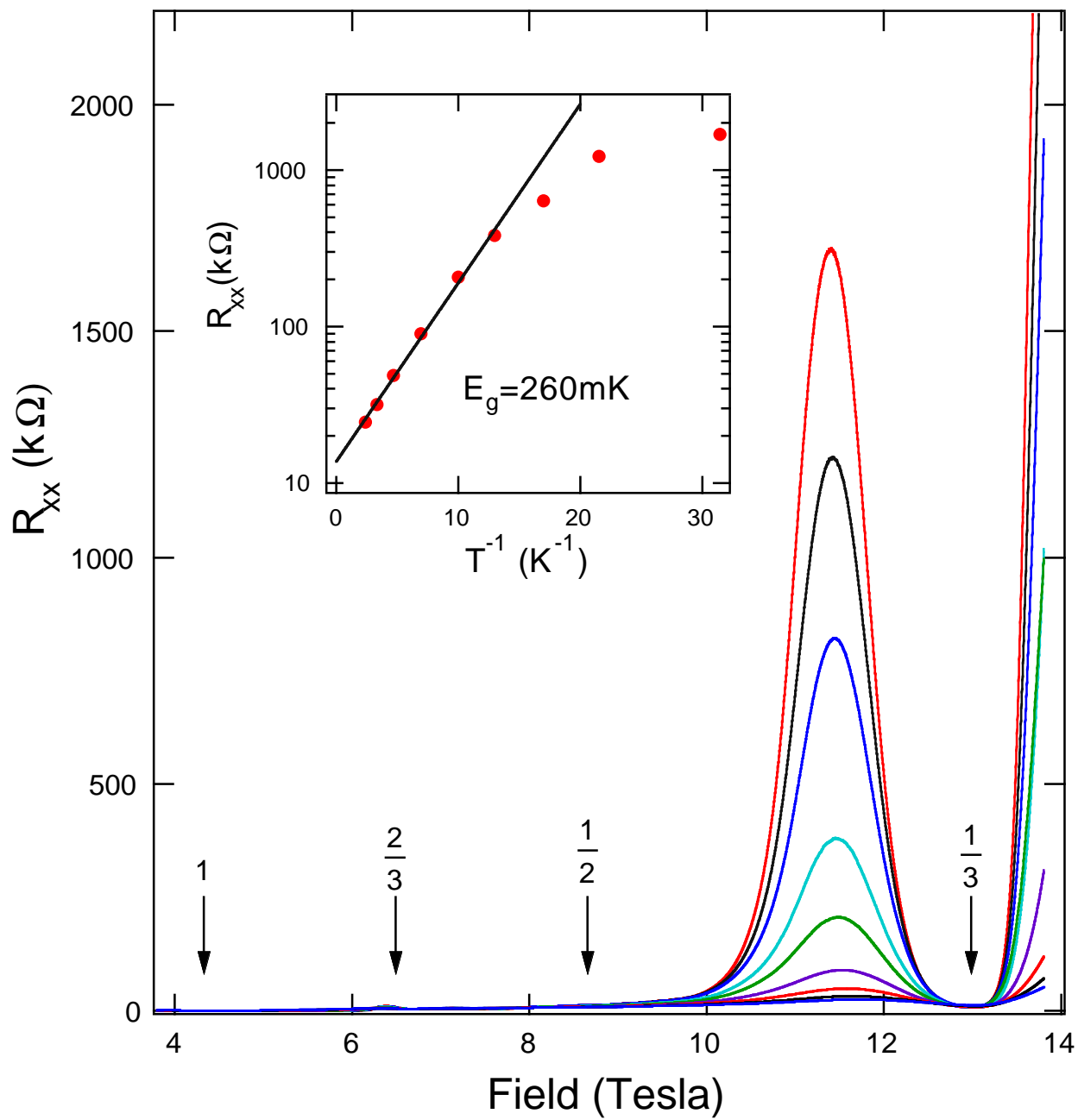
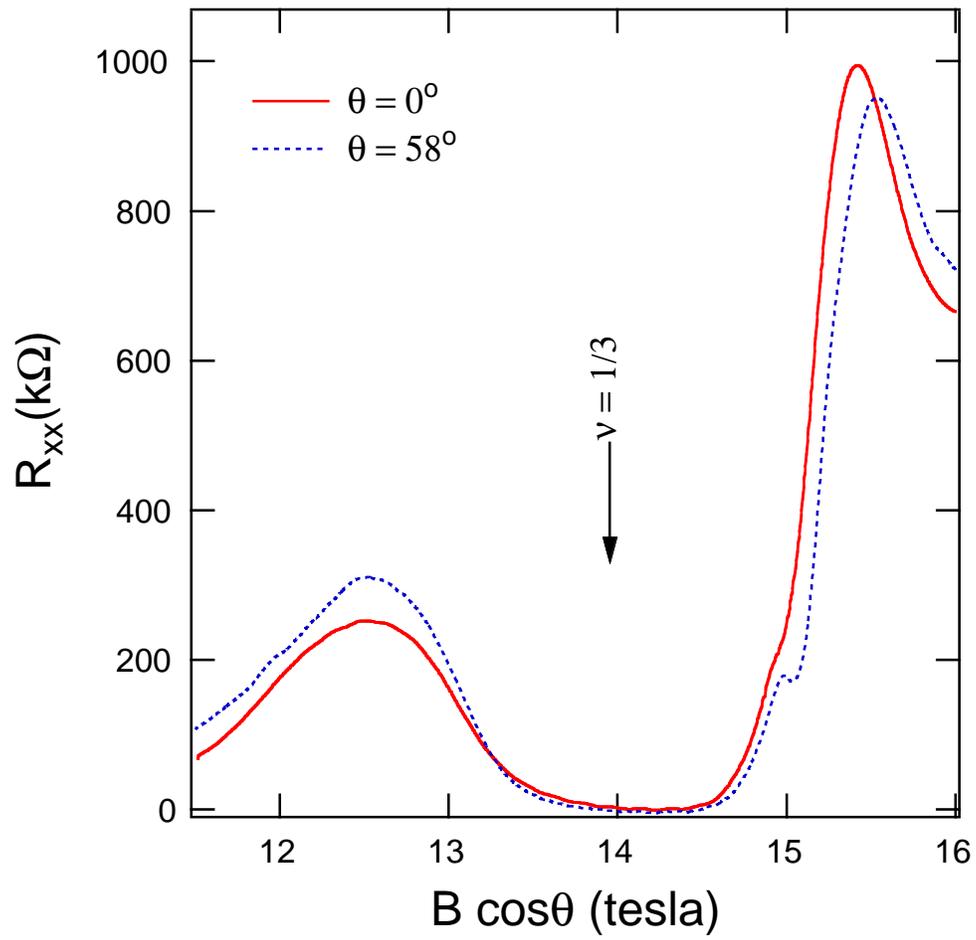
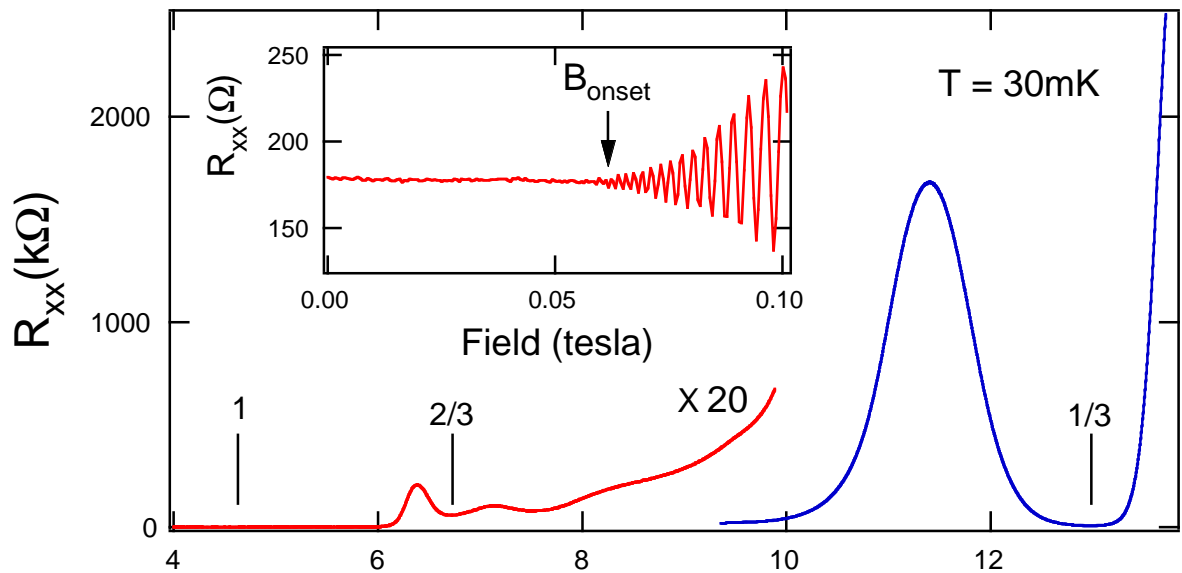
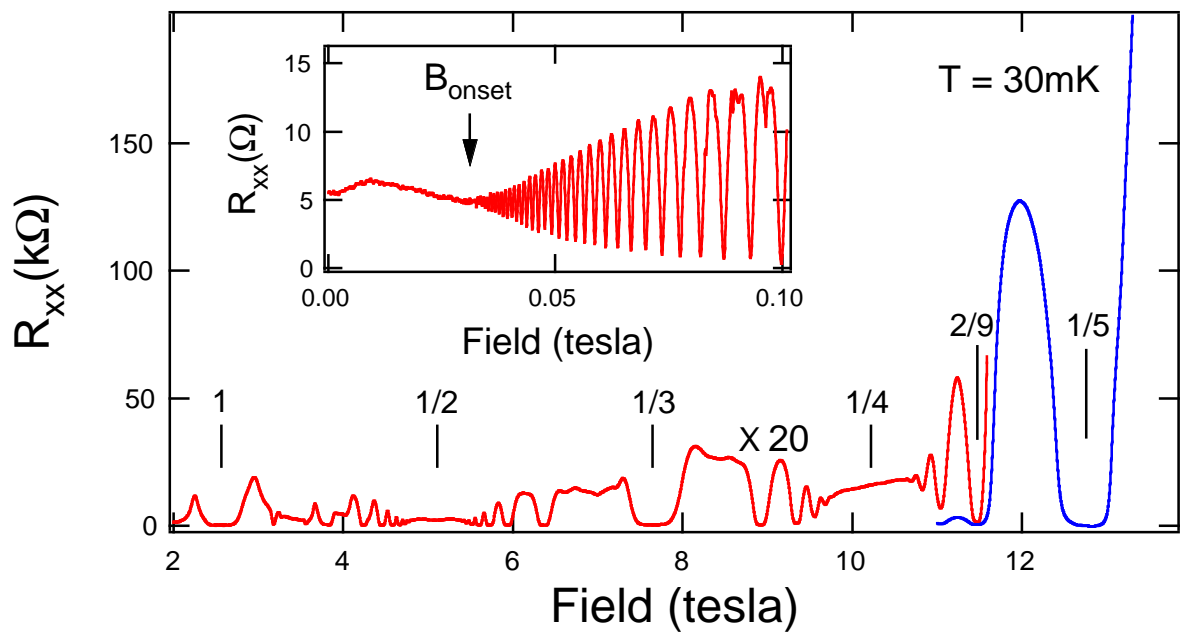
\end{document}